\documentclass[12pt,a4paper,twoside, twocolumn]{article}
\usepackage{dhasa}
\usepackage{caption}
\usepackage{subcaption}
\begin{document}

\title*{An Evaluation of GPT-4V for Transcribing the Urban Renewal Hand-Written Collection}
\textit{Lee, Myeong}\\
\textit{mlee89@gmu.edu}\\
\textit{George Mason University, Fairfax, U.S.} 

\textit{Hsu, Julia H. P.}\\
\textit{hhsu2@gmu.edu}\\
\textit{George Mason University, Fairfax, U.S.}

\section{Introduction}

From the early 1960s to the mid-1970s, Asheville, North Carolina, underwent urban renewal, a national program aimed at modernizing "blighted" areas~\citep{lee2017heuristics}. This process, mostly impacting African-American neighborhoods, displaced families, businesses, and organizations for economic and infrastructure development.
Due to its historical significance and on-going reparation efforts in many cities in the United States, understanding parcel-level information about how each property was acquired by the housing authority is one of the most important steps. 
This paper focuses on the East Riverside neighborhood (now Southside), the southeastern U.S.'s largest area affected by urban renewal. 
Our target documents are property acquisition documents originally produced by the Housing Authority of the City of Asheville (HACA) that details the property acquisition processes of nearly 1,000 properties in Southside. They include property's complete sales history, including appraisals, offers, rejections, and court cases.

However, the extensive lengths (20-200 pages per parcel) and complexity of property acquisition documents from the urban renewal era pose significant challenges in digital curation, particularly for handwritten documents. This is a persistent issue for computational archival scientists and digital humanities scholars.
Although handwritten text recognition (HTR) platforms such as Transkribus \citep{kahle2017transkribus} and Google Document AI\footnote{https://cloud.google.com/document-ai} have been extensively studied with good performances (e.g., \citep{ingle2019scalable,kiessling2019escriptorium}), their practical application by non-experts remains challenging. This is due to the additional technical efforts required to configure recognition systems, such as training for specific document types, data structuring, and cleaning processes.

In November 2023, OpenAI released GPT-4V(ision), which includes Optical Character Recognition (OCR) capabilities. Given that much of the data curation, processing, and cleaning can be managed through user-friendly prompts (i.e., chat), we aim to conduct an initial assessment of GPT-4V's effectiveness in transcribing hand-written documents from the urban renewal collection.
If GPT-4V can accurately digitize hand-written documents through carefully crafted prompts, it could become a valuable tool for non-experts in transcribing historical documents on a large scale. Alternatively, if it falls short, it is still crucial to understand and discuss the implications of using Large Language Models (LLMs) for digitizing archival documents.

\begin{figure*}[h!]
    \centering
    \includegraphics[width = 0.85\textwidth]{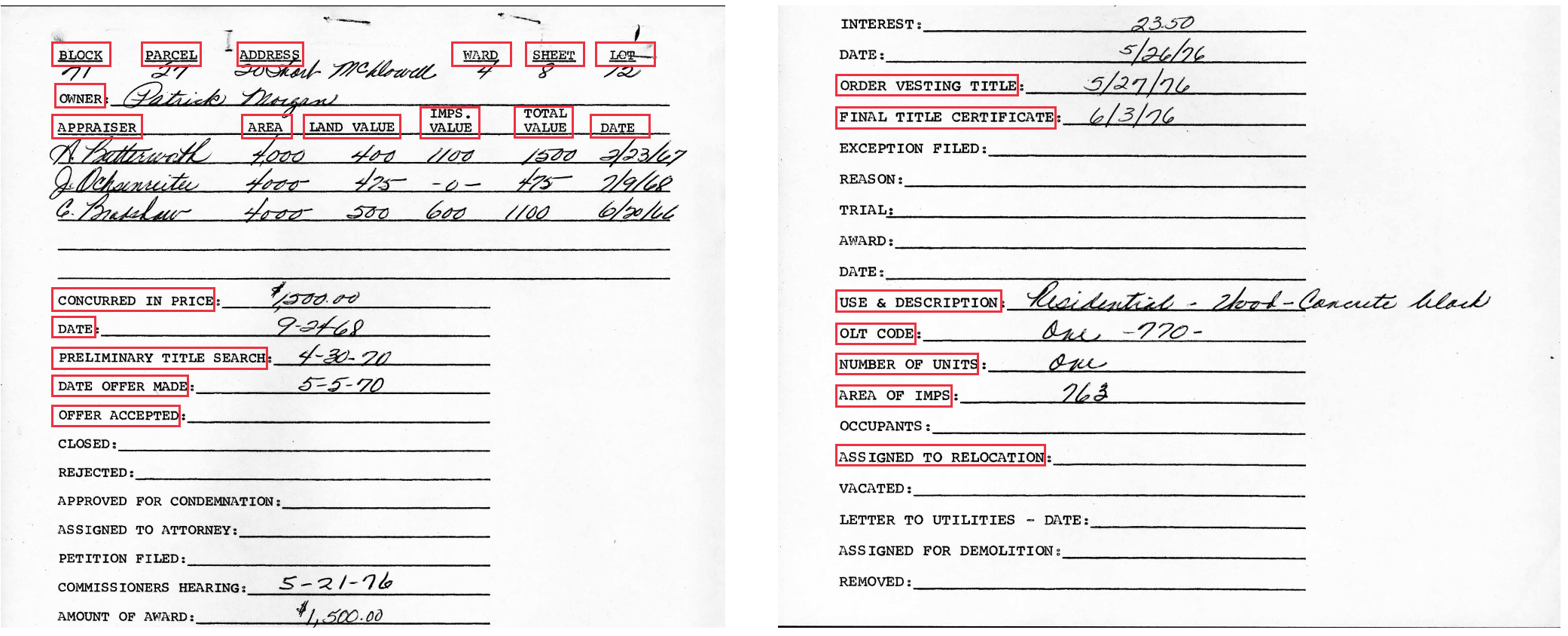}
     \caption{An example of original document. Red rectangles are fields of our focus in transcription.}
     \label{fig:org_doc}
\end{figure*}

This paper evaluates GPT-4V's performance in transcribing the cover pages of selected urban renewal documents. These cover pages, all hand-written and generally more challenging to read (even for humans) compared to other parts of the documents, are valuable for researchers and practitioners focusing on urban renewal, as they succinctly provide key information about property acquisition processes. Figure~\ref{fig:org_doc} shows an example of an original document.

\section{Approach}
\begin{figure}[b!]
\centering
     \includegraphics[width=\columnwidth]{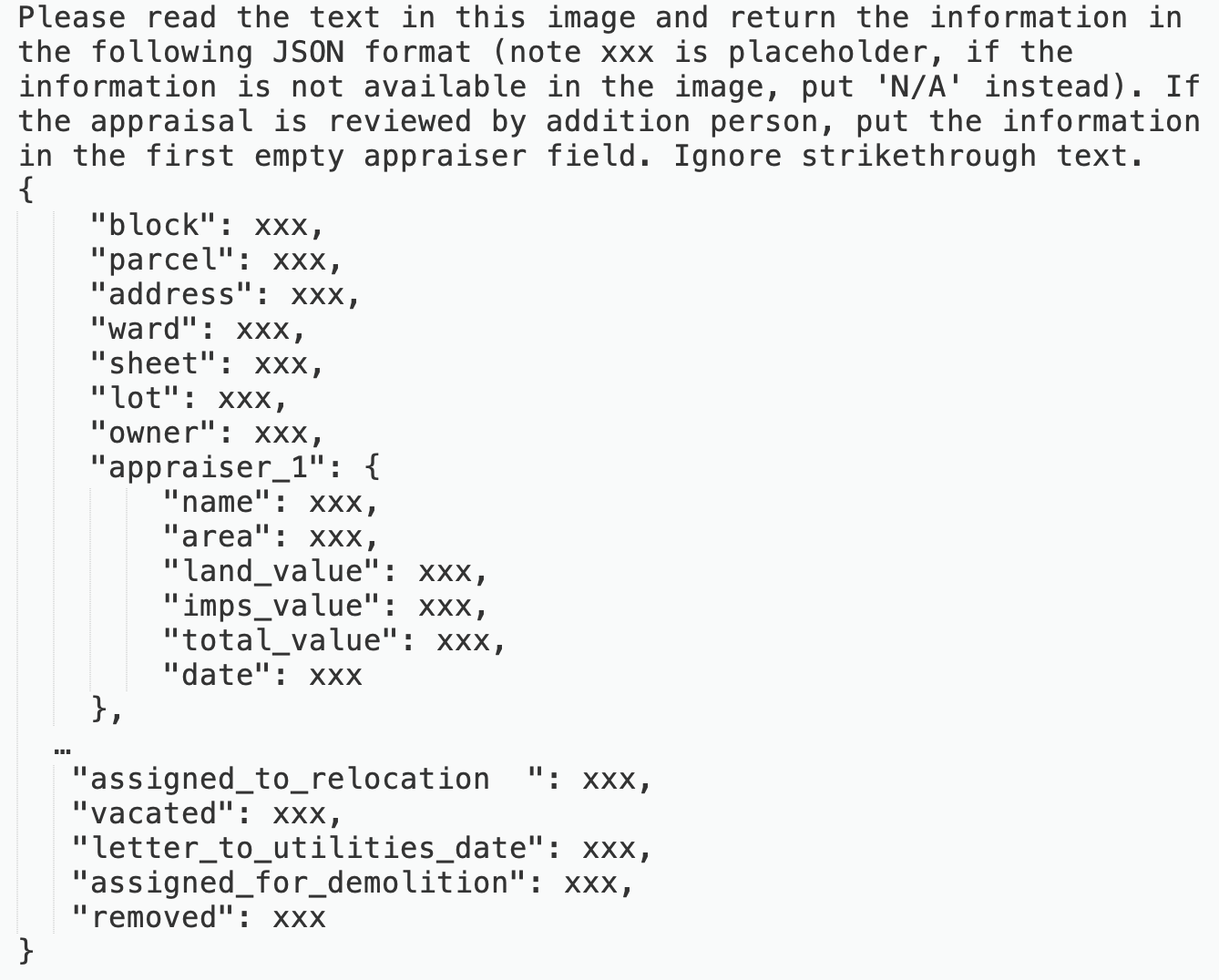}
     \caption{Constrained Prompting}
     \label{fig:constrained_prompt}
\end{figure}

We utilized the OpenAI APIs \footnote{\url{https://platform.openai.com/docs/overview}} to make queries to GPT-4V. We employed constrained prompting techniques to instruct GPT-4V to read the text in the document~\citep{yang2023dawn}. The objective was to retrieve information from individual fields and organize it into a specific JSON format. 
We engineered the prompt following the guideline by \cite{yang2023dawn}. Through iterative refinements of the prompts, the constrained prompt was finalized as depicted in Figure~\ref{fig:constrained_prompt}. 
Using this prompt, we processed $50$ select documents, subsequently assessing its performance per each field. 

This process involved comparing the text within the retrieved JSON data by GPT-4V with the ground truth data compiled by human evaluators. Given that several fields were frequently empty across most documents (e.g., Exception filed, Trial), our focus was on fields with values present in more than $50\%$ of the samples ($25$ documents). It is worth noting that the accuracy was calculated solely based on non-empty values.
Based on the accuracy of each field's transcribed values, we qualitatively examined the characteristics of the hand-written letters to understand the strengths and weaknesses of GPT-4V. 

% , while the other represented the overall accuracy regardless whether the value was empty or not.

\section{Results}
% Figure \ref{fig:returned_json} shows an example of JSON data returned by GPT-4Vision, where incorrect answers were highlighted in red.
The processing time for GPT-4V to read $50$ documents was about $19.11$ minutes.
Table \ref{tab:accuracy} presents the accuracy for each individual field. The \textit{Accuracy} column of Table \ref{tab:accuracy} reports the accuracy solely based on non-empty values. Notably, the field \textit{ward} achieved the highest accuracy score of $0.8$, whereas \textit{owner} obtained the lowest accuracy score of $0.18$ in this evaluation.
Our initial assessment suggests that, without further engineering of prompts and additional frameworks that help refine the results, GPT-4V is not at the stage of generalizable transcription tasks for hand-written texts. 

To understand the reasons for the varying accuracy in transcribing tasks, we qualitatively analyzed the hand-written texts by comparing between different fields. 
Notably, fields that present high accuracy such as \textit{ward} and \textit{block} are more likely to provide numerical values in relatively simple forms, providing clear structures for comprehension by GPT-4V. Among the fields with more than 0.7 of accuracy are all simple numeric fields or dates, with an exception of the \textit{use\_description} field. 
Although the \textit{use\_description} field is English words, it presents relatively standardized terms, such as "Residential" or "Commercial", which may made it easy for GPT-4V to predict the values. 

Conversely, fields with low accuracy, such as \textit{owner} and \textit{appraiser\_$1$\_name}, are attributed to their longer, more diverse text entries, making it challenging for GPT-4V to extract specific information accurately. 
Finally, we identified consistent misreads of GPT-4V. They include, but are not limited to, a confusion between "$2$" and "$7$," "$1$" and "$7$," "$0$" and "$6$," and "$F$" and "$E$". These recurring errors illustrate consistent challenges faced by the model in distinguishing specific numerical and alphabetical characters accurately.

\begin{table*}[t]
\caption{Accuracy of each field in the cover page of the urban renewal collection.}
\centering
\begin{tabular}{|l|r|r|}
\hline
Field & Accuracy & \# empty cells\\
\hline
ward & $0.80$ & $0$\\
block & $0.76$ & $0$\\
appraiser\_$1$\_land\_value & $0.74$ & $0$ \\
order\_vesting\_title & $0.74$ & $8$ \\
use\_description & $0.74$ & $8$ \\
sheet & $0.72$ & $0$\\
appraiser\_$1$\_total\_value & $0.72$ & $0$\\
date\_offer\_made & $0.69$ & $8$ \\
offer\_accepted & $0.68$ & $25$ \\
final\_title\_certificate & $0.67$ & $8$ \\
appraiser\_$2$\_total\_value & $0.63$ & $1$ \\
preliminary\_title\_search & $0.63$ & $4$ \\
parcel & $0.62$ & $0$\\
appraiser\_$2$\_land\_value & $0.61$ & $1$ \\
appraiser\_$1$\_imps\_value & $0.60$ & $25$ \\
appraiser\_$1$\_date & $0.59$ & $1$ \\
concurred\_in\_price\_date & $0.59$ & $6$ \\
assigned\_to\_relocation & $0.58$ & $14$ \\
appraiser\_$3$\_date & $0.57$ & $8$ \\
appraiser\_$3$\_total\_value & $0.57$ & $6$ \\
lot & $0.56$ & $0$\\
concurred\_in\_price & $0.54$ & $2$ \\
appraiser\_$1$\_area & $0.54$ & $0$ \\
appraiser\_$2$\_date & $0.53$ & $1$ \\
appraiser\_$2$\_area & $0.51$ & $1$ \\
appraiser\_$3$\_name & $0.41$ & $6$ \\
number\_of\_units & $0.41$ & $23$ \\
appraiser\_$2$\_name & $0.39$ & $1$ \\
olt\_code & $0.31$ & $15$ \\
address & $0.30$ & $0$ \\
area\_of\_imps & $0.29$ & $22$ \\
appraiser\_$1$\_name & $0.22$ & $0$ \\
owner & $0.18$ & $0$ \\
\hline
\end{tabular}
\label{tab:accuracy}
\end{table*}

\section{Discussion}
Our findings offer valuable insights for using LLMs like GPT-4V to transcribe handwritten documents, particularly historical ones. Because LLMs exhibit varied performance in different types of texts, further examination is necessary for their application in transcribing handwritten texts. However, they may be suitable for simple tasks, as they tend to better recognize numerical and basic textual content.

Our research suggests several potential directions for future research. First, since GPT-4V commonly misreads certain characters (e.g., confusing 1 and 7), these patterns could be integrated into prompt design in a recursive manner. Second, for text-based fields, options could be recursively narrowed down through trial and error, potentially incorporating human oversight. Lastly, considering the diverse characteristics and performances of various LLMs across different handwriting styles, an ensemble approach combining multiple LLM tools could be developed to enhance transcription accuracy.

% \begin{figure}[b!]
% \includegraphics[width=\columnwidth]{SingleFileStyleGuide-img002.png}
% \caption{Example of a column-width figure\\ Mikeal Coxous: Bandeau du Manifeste des Digital Humanities, public domain}
% \end{figure}

\subsection*{Acknowledgments}
We thank Ms. Priscilla Robinson and Dr. Richard Marciano for their continuing collaboration and the scanning of the archival documents. 

%%DO NOT DELETE
\bibliographystyle{agsm}

%%Provide your own references in the bibliography.bib included in this guide %% 
\bibliography{bibliography.bib}

\end{document}